\documentclass[12pt]{article}

\usepackage{scicite}
\usepackage{times}

\usepackage{graphicx} 
\usepackage{subfig}
\usepackage[articletitle=true,style=science,maxbibnames=99]{biblatex}
\bibliography{main.bib}

\usepackage{url}
\usepackage{comment}

\usepackage{setspace}

\usepackage{amsmath}

\usepackage{xcolor}

\newcommand{\rzero}{$R_0$\;}

\newenvironment{sciabstract}{%
\begin{quote} \bf}
{\end{quote}}

\title{Worldwide wildfire spreading and its severity described by the SIR model}

\author
{Tong Pan$^{1\dagger}$, Hongjun Wang$^{2\dagger}$, Jiyuan Chen$^{3}$, \\ Xuan Song$^{1\ast}$,\\
\\
\normalsize{$^{1}$Department of Computer Science and Engineering,}\\ 
\normalsize{Southern University of Science and Technology, Shenzhen, Guangdong, China}\\
\normalsize{$^{2}$Hong Kong Polytechnic University, Hong Kong SAR, China}\\
\\
\normalsize{$^\ast$To whom correspondence should be addressed;} \\
\normalsize{E-mail: 1155129239@link.cuhk.edu.hk.}
}

\date{}

\begin{document} 

\baselineskip24pt

\maketitle


\begin{sciabstract}

Abstract:
\end{sciabstract}


\noindent \textbf{One-Sentence Summary:} 


\section*{Introduction}
\section*{Results}
\subsection*{Fire dynamics modeling}
\begin{table*}
	\centering
	\begin{tabular}{cc}
		\hline
		\hline
		Continent & Number of fire events \\
		\hline
		Africa & 23479770\\
		South America & 6190501 \\
		Asia & 6090517 \\
		North America & 2324829\\
		Europe & 2324584 \\
		Oceania & 1192404 \\
		Seven Seas & 1411 \\
		\hline
		Global & 41789913\\
		\hline
		\hline
	\end{tabular}
	\caption{Number of fire events from year 2002 to 2023 at continent level. }
	\label{tab:total_num_fire_continent}
\end{table*}

\begin{table*}
	\centering
	\begin{tabular}{cc}
		\hline
		\hline
		Continent & Burned Area (million km$^2$) \\
		\hline
		Africa & 96.34 \\
		South America & 26.41\\
		Asia & 20.07 \\
		Europe & 10.78 \\
		Oceania & 9.84 \\		
		North America & 9.26\\
		Seven Seas & 3.00  \\
		\hline
		Global & 173.32 \\
		\hline
		\hline
	\end{tabular}
	\caption{Burned Area in million square kilometers from year 2002 to 2023 at continent level. }
	\label{tab:BA_continent}
\end{table*}

\begin{table*}
	\centering
	\begin{tabular}{cc}
		\hline
		\hline
		Top-10 Countries & Number of fire events \\
		\hline
		Democratic Republic of the Congo & 4748743\\
		Brazil & 3516396 \\
		Angola & 2934788 \\
		Zambia & 2090508 \\
		Central African Republic & 1554983 \\
		Russia & 1545821\\
		Mozambique & 1517531 \\
		People's Republic of China & 1241056\\
		South Sudan & 1200230\\
		Australia & 1073981\\
		\hline
		\hline
	\end{tabular}
	\caption{Top 10 countries with the total number of fire events. }
	\label{tab:total_num_fire_country}
\end{table*}

\begin{table*}
	\centering
	\begin{tabular}{cc}
		\hline
		\hline
		Top-10 Countries & Burned Area (million km$^2$) \\
		\hline
		Democratic Republic of the Congo & 18.52 \\
		Brazil & 15.43\\
		Angola & 14.38 \\
		Australia & 9.51 \\
		Zambia & 8.43\\
		Russia & 8.33 \\		
		South Sudan & 7.06\\
		Central African Republic & 6.45\\
		Mozambique & 5.83\\
		United State of America & 3.80\\
		\hline
		\hline
	\end{tabular}
	\caption{Top 10 countries with Burned Area. }
	\label{tab:BA_country}
\end{table*}

\begin{figure*}[t] 
	\centering
	\captionsetup[subfigure]{} 
	\subfloat[2019-2020 Australian Bushfire Season]{\label{} \includegraphics[width=0.6\textwidth]{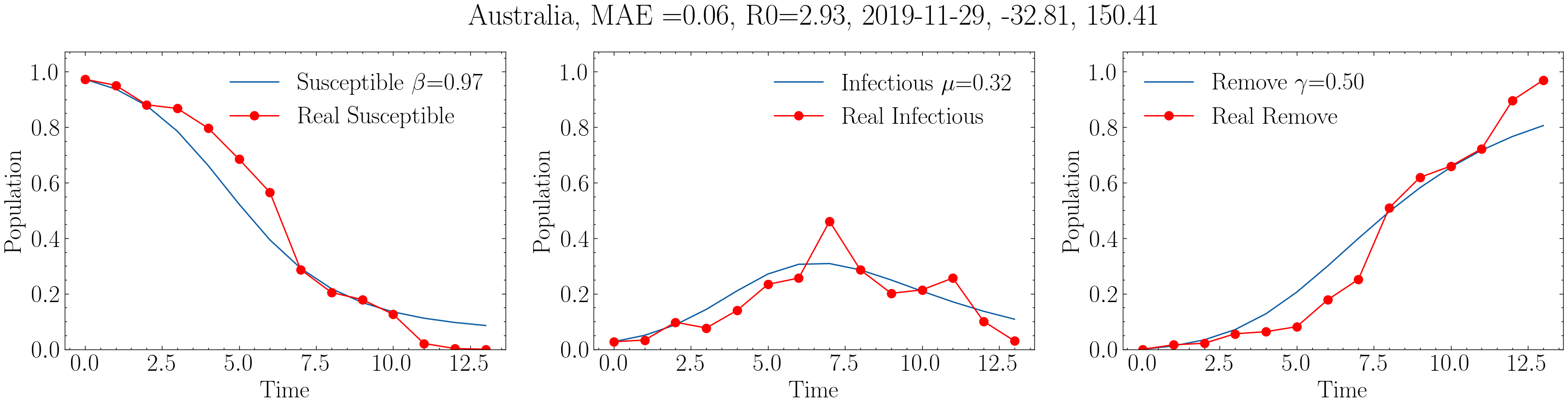}}\\
	\subfloat[2019-2020 Australian Bushfire Season]{\label{} \includegraphics[width=0.6\textwidth]{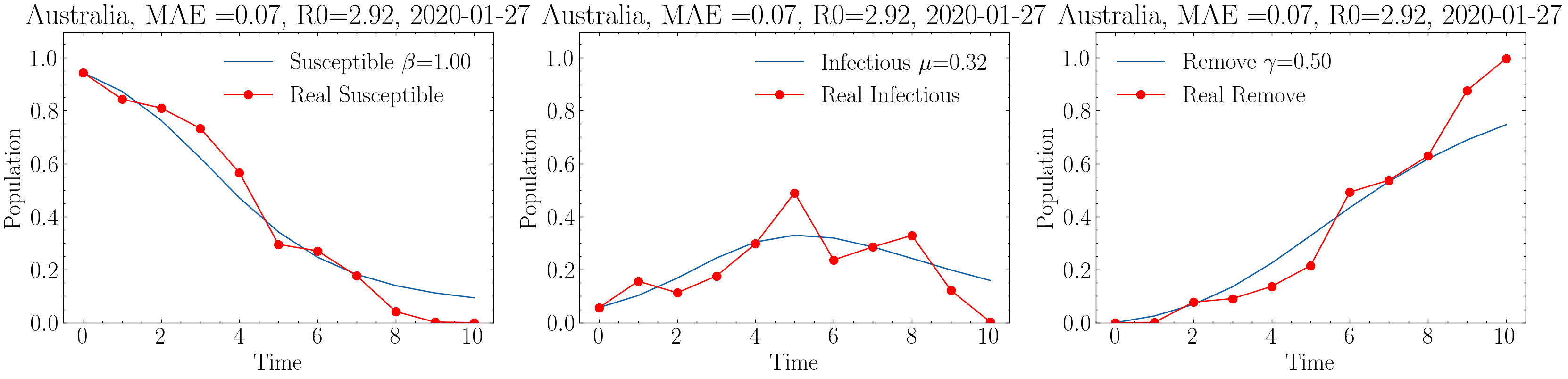}}\\
	\subfloat[Bearskin Fire in Idaho]{\label{} \includegraphics[width=0.6\textwidth]{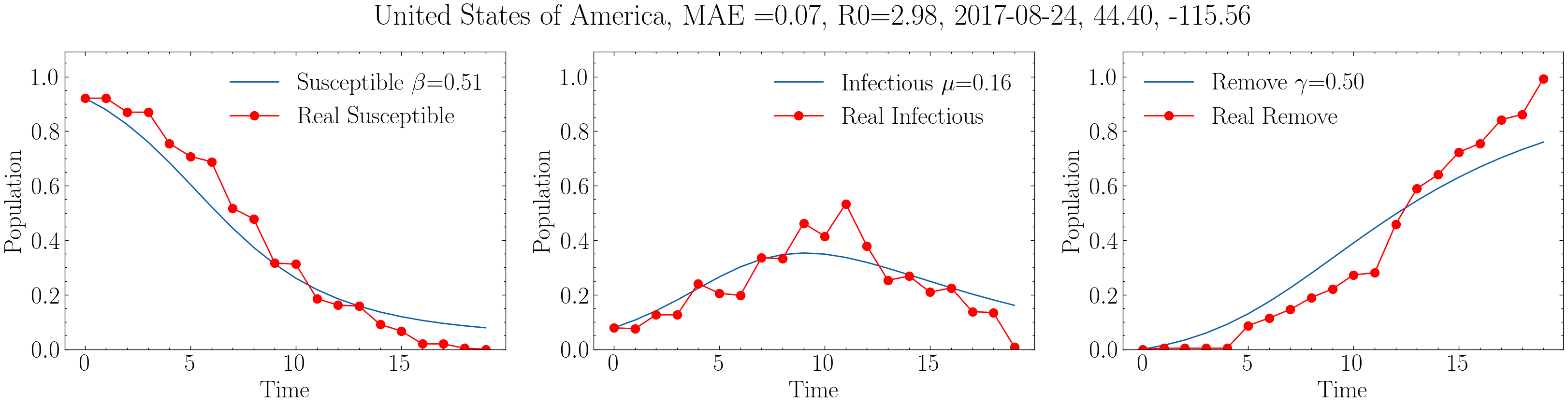}}\\
	\subfloat[Southern California wildfires]{\label{} \includegraphics[width=0.6\textwidth]{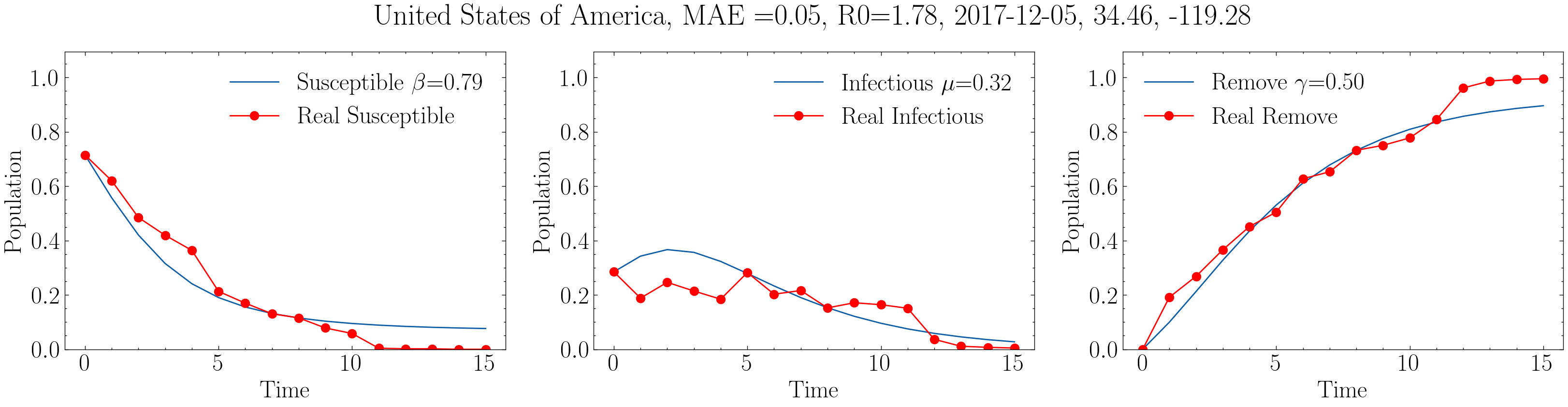}}\\
	\caption{SIR dynamics for a fire event in Australia and the United State of America, respectively. }
	\label{fig:SIR_4events}
\end{figure*}

We estimate that from 2002 to 2020, there are in total 36527974 recorded wildfires around the world, near two million fire events per year. At the continent level, we summarized the total number of recorded wildfires and the Burned Area (BA) for each continent in Table~\ref{tab:total_num_fire_continent} and \ref{tab:BA_continent}, where Africa (20392837, 84.20 million km$^2$) was found to to have highest contribution to the number of wildfires, followed by South America (5435623, 23.24 million km$^2$), Asia (5335222, 17.87 million km$^2$) and Europe (2151347, 9.83 million km$^2$). At the country level, as shown in Table~\ref{tab:total_num_fire_country}, Democratic Republic of the Congo shares the largest number of wildfires in the past two decades, followed by Brazil, Angola and Zambia. For the country ranking of BA (Table~\ref{tab:BA_country}), Democratic Republic of the Congo is also the top country, followed by Brazil, Angola and Australia. 

Among the fire events, every year worldwide some extraordinary wildfires occur, overwhelming suppression capabilities, causing substantial damages, and often resulting in fatalities \cite{fire1010009}. For example, extreme, long-duration wildfires in the western USA in 2017 \cite{fire1010018}, and in eastern Australia in 2019-2020 \cite{boer2020unprecedented} had attracted social attention. Here, we take the above-mentioned fire events as examples to demonstrate the dynamics of fire events and provide the metric of the fire spread (Fig.~\ref{fig:SIR_4events}). The dynamics of fire events can be well-modeled by the susceptible-infected-recovered (SIR) model, details of which will be shown in Data and methods. The observed number of spatio-temporal fire grids for each compartment (compartment $S$: left subplot; compartment $I$: middle subplot; compartment $R$: right subplot) can be well described with the SIR model, with the Mean Absolute Errors (MAEs) are (a) 0.06; (b) 0.07; (c) 0.07; (d) 0.05. The basic reproductive number, \rzero, which is calculated per event, represents the average number of secondary fire grids ignited by a primary fire event, serving as a key metric of its potential to spread rapidly. For extreme fire events in Fig.~\ref{fig:SIR_4events}, \rzero are (a) 2.93 (b) 2.92 (c) 2.98 and (d)1.78. Fitting empirical data with SIR model, a set of \rzero for all events of interest can be obtained. 

\subsubsection*{Global distribution of averaged \rzero during 2002-2023}
\begin{figure*}[htbp] 
	\centering
	\includegraphics[width=0.6\textwidth]{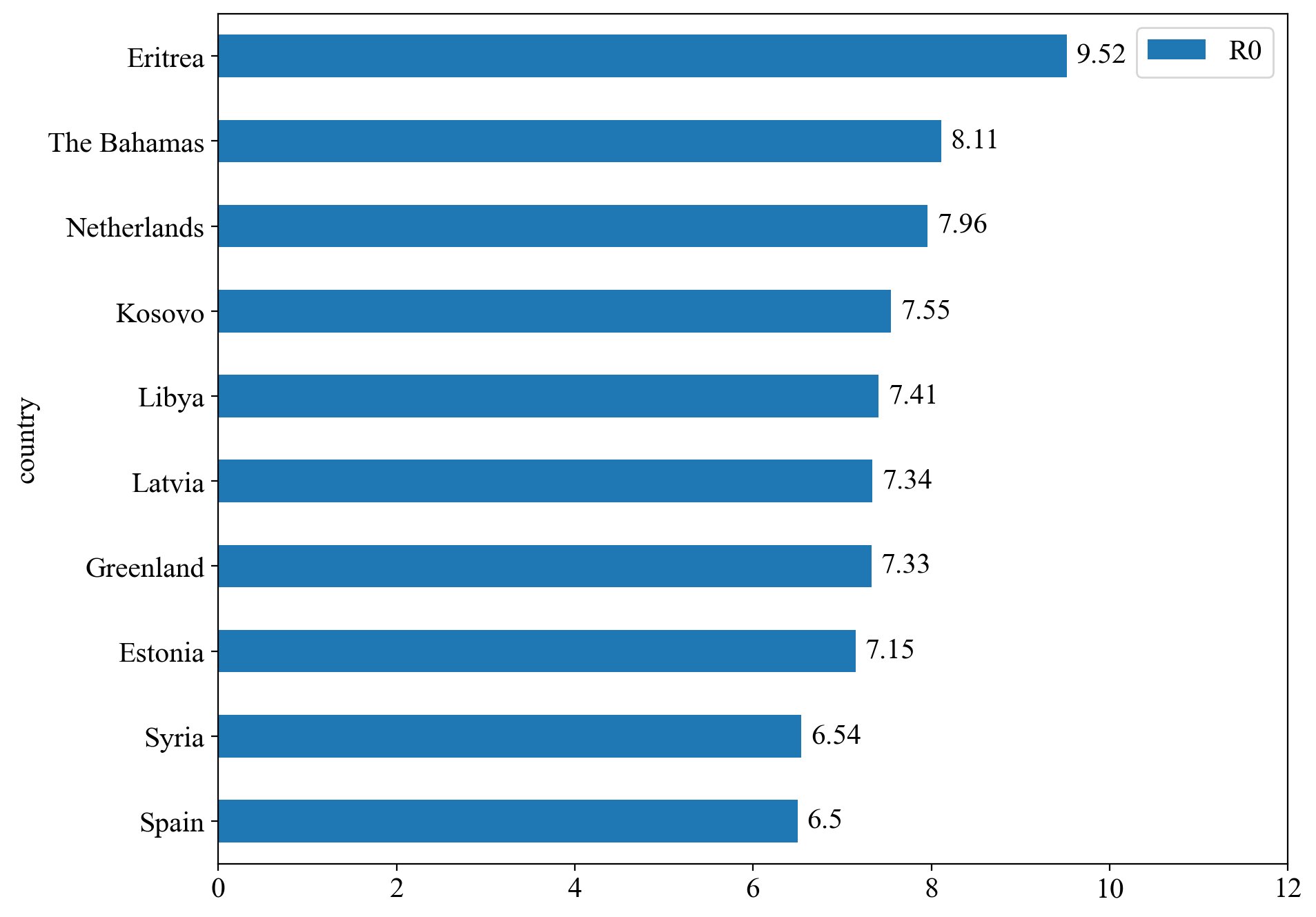}\\
	\caption{Top 10 countries for \rzero.}
	\label{fig:R0_global_ranking_Top10}
\end{figure*}

\begin{figure*}[htbp] 
	\centering
	\includegraphics[width=\textwidth]{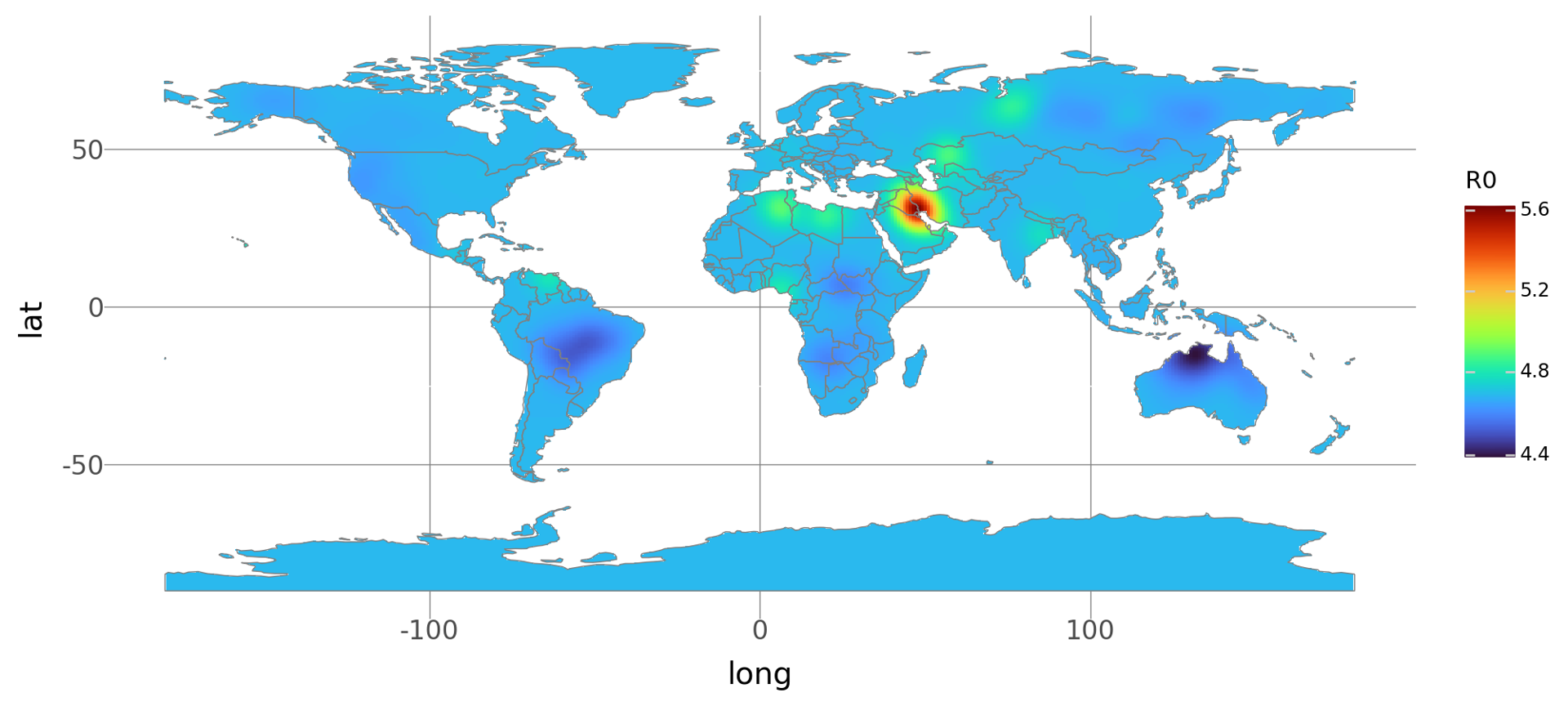}\\
	\caption{Global distribution of averaged \rzero over two decades.}
	\label{fig:R0_global}
\end{figure*}

\textcolor{red}{(We found a more clustered distribution of \rzero. )} We found that there exists obvious heterogeneity in the global distribution of averaged $R_0$ per event during 2002-2020, as shown in Fig.~\ref{fig:R0_global}. Middle East, North Africa, Northeast India, Kazakhstan, and Central Russia share relatively high value of $R_0$ (greater than 5). At the continent level, as shown in Table~\ref{tab:R0_continent}, Asia takes the lead with an averaged \rzero  of 5.90, followed by Africa (4.68), Europe (4.54) and North America (4.11). At the country level, Eritrea (9.52), The Bahamas (8.11), Netherlands (7.96) are the ranked Top-3 \rzero countries as shown in the ranking plot Fig.~\ref{fig:R0_global_ranking_Top10} and Table~\ref{tab:R0_country}, followed by Kosovo (7.55) and Libya (7.41). 
\begin{table*}
	\centering
	\begin{tabular}{ccc}
		\hline
		\hline
		Continent & \rzero & MAE\\
		\hline
		Asia & 5.90 & 0.09\\
		Africa & 4.68 & 0.10\\
		Europe & 4.54 & 0.09\\
		North America & 4.11 & 0.09\\
		South America & 4.00 & 0.09\\
		Oceania & 2.87 & 0.11\\
		Seven Seas & 2.78 & 0.12\\
		
		\hline
		Global & 4.79 & 0.10 \\
		\hline
		\hline
	\end{tabular}
	\caption{}
	\label{tab:R0_continent}
\end{table*}

\begin{table*}
	\centering
	\begin{tabular}{ccc}
		\hline
		\hline
		Top-10 Countries & \rzero & MAE\\
		\hline
		Eritrea & 9.52 & 0.06\\
		The Bahamas & 8.11 & 0.05\\
		Netherlands & 7.96 & 0.12\\
		Kosovo & 7.55 & 0.20\\
		Libya & 7.41 & 0.09\\
		Latvia & 7.34 & 0.09\\
		Greenland & 7.33 & 0.08\\
		Estonia & 7.15 & 0.10 \\
		Syria & 6.54 & 0.10\\
		Spain & 6.50 & 0.11\\
		\hline
		\hline
	\end{tabular}
	\caption{}
	\label{tab:R0_country}
\end{table*}

\begin{figure*}[htbp] 
	\centering
	\includegraphics[width=0.5\textwidth]{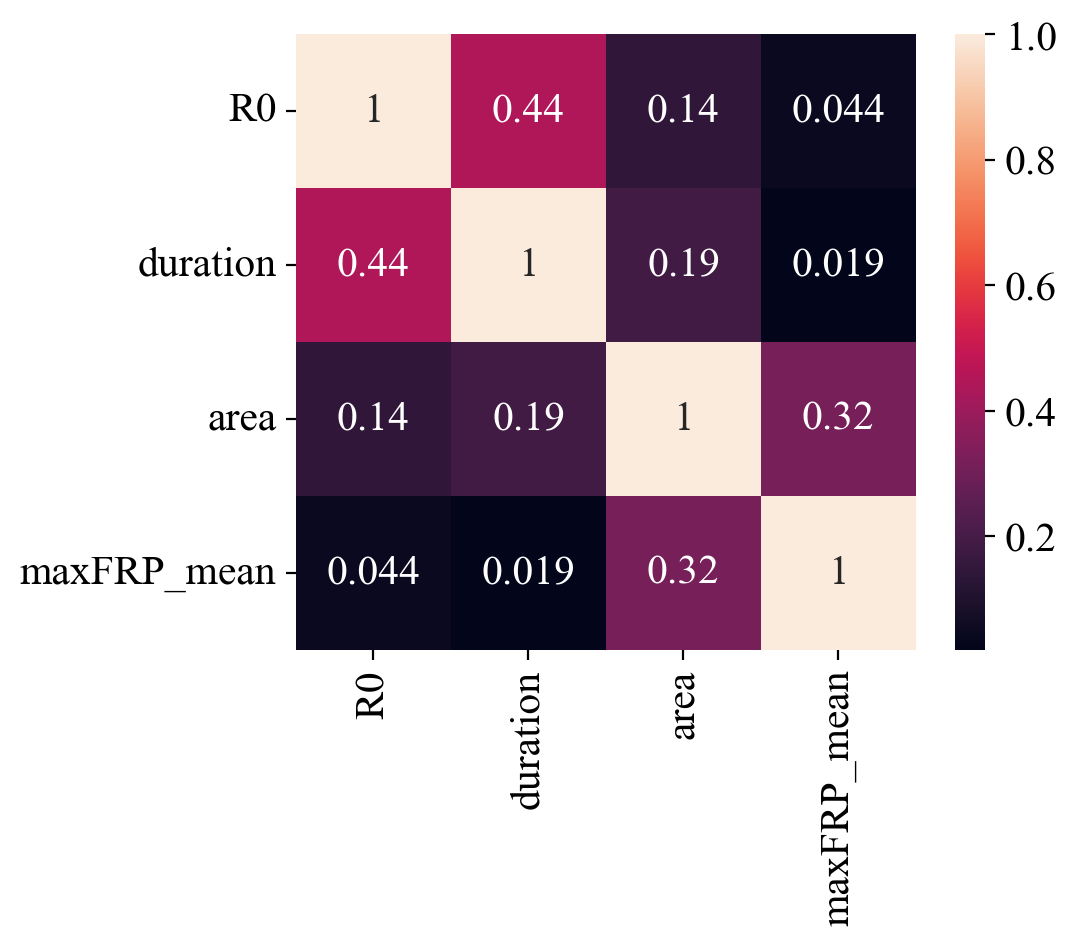}\\
	\caption{Correlation matrix between \rzero, duration, Burned Area (BA) and maximum Fire Radiative power (max FRP). }
	\label{fig:R0_correlation_matrix}
\end{figure*}

\subsubsection*{Historical evolution of global \rzero during 2002-2023}
\begin{figure*}[htbp] 
	\centering
	\subfloat[Africa]{\label{fig:R0_Africa_temporal} \includegraphics[width=0.45\textwidth]{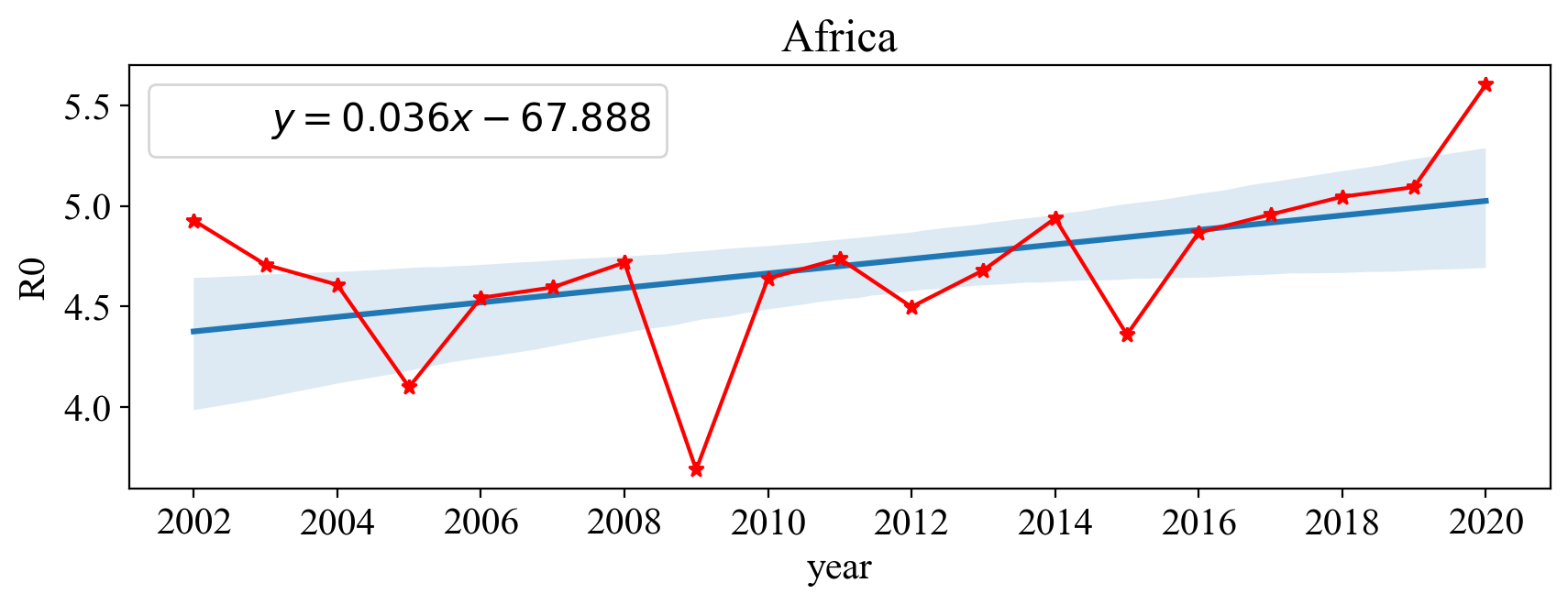}}
	\subfloat[Asia]{\label{} \includegraphics[width=0.45\textwidth]{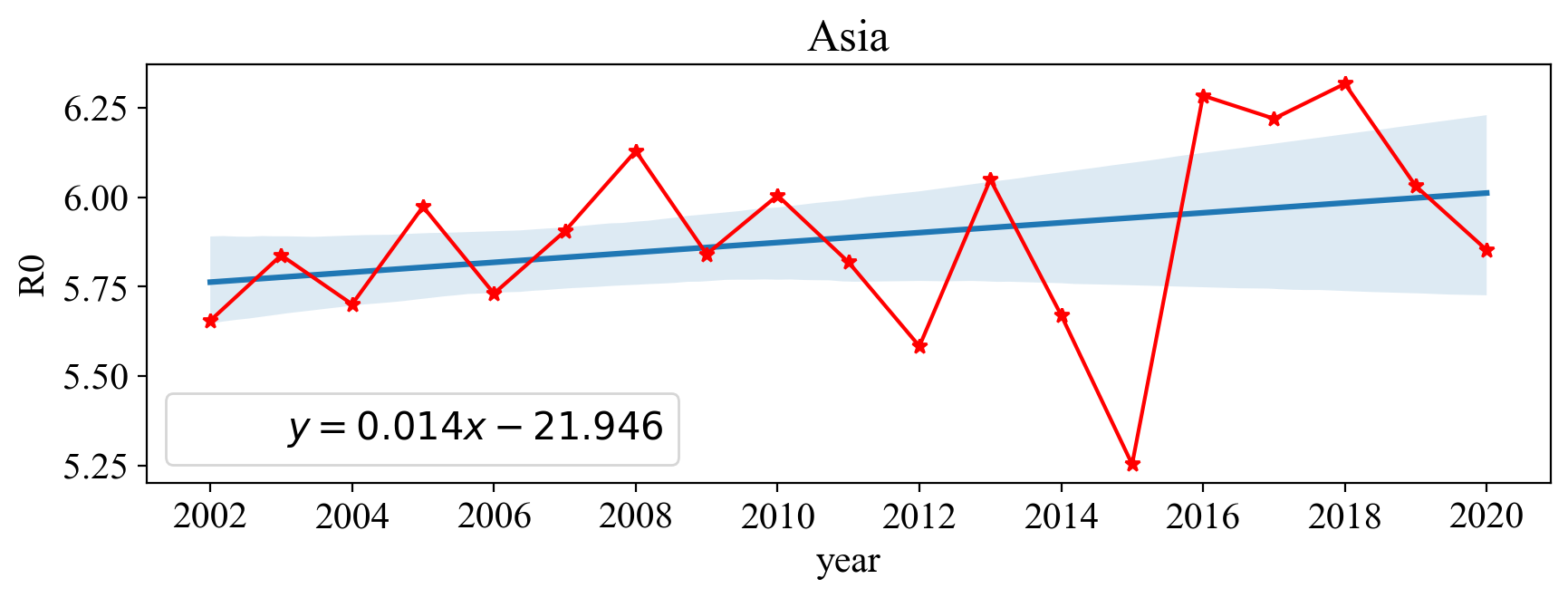}}\\
	\subfloat[Europe]{\label{} \includegraphics[width=0.45\textwidth]{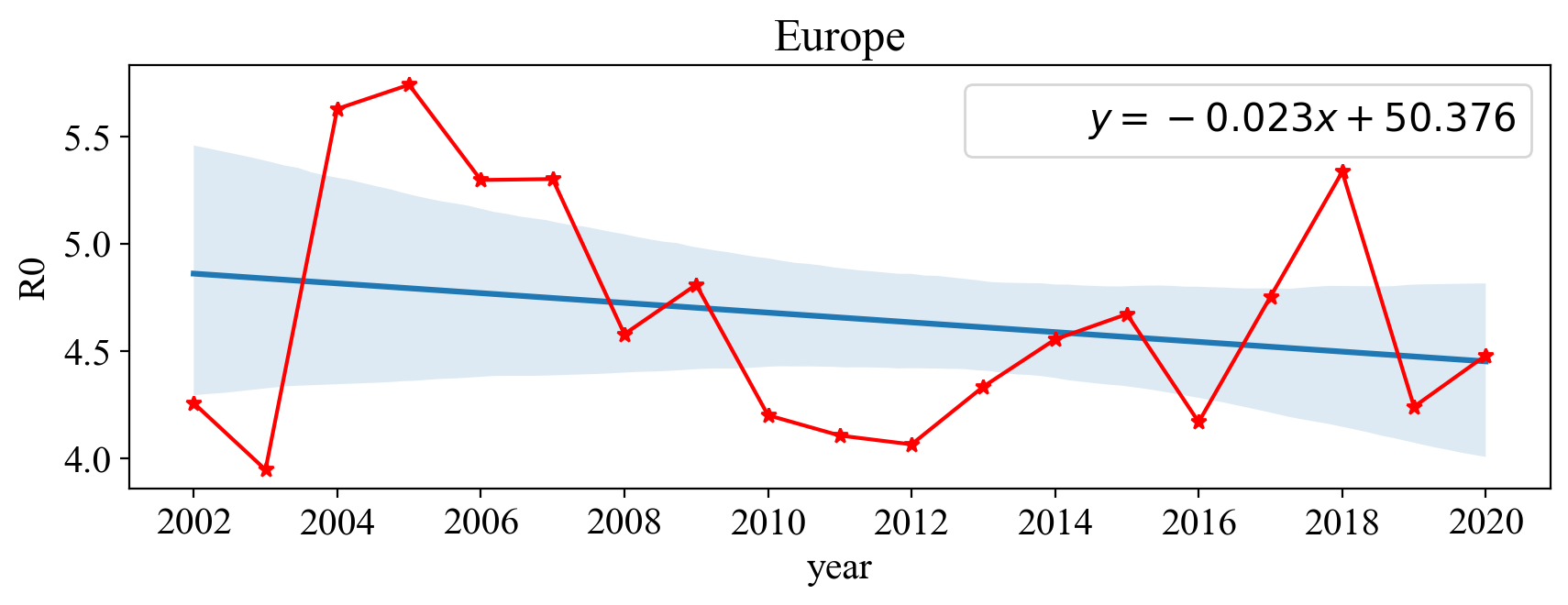}}
	\subfloat[North America]{\label{} \includegraphics[width=0.45\textwidth]{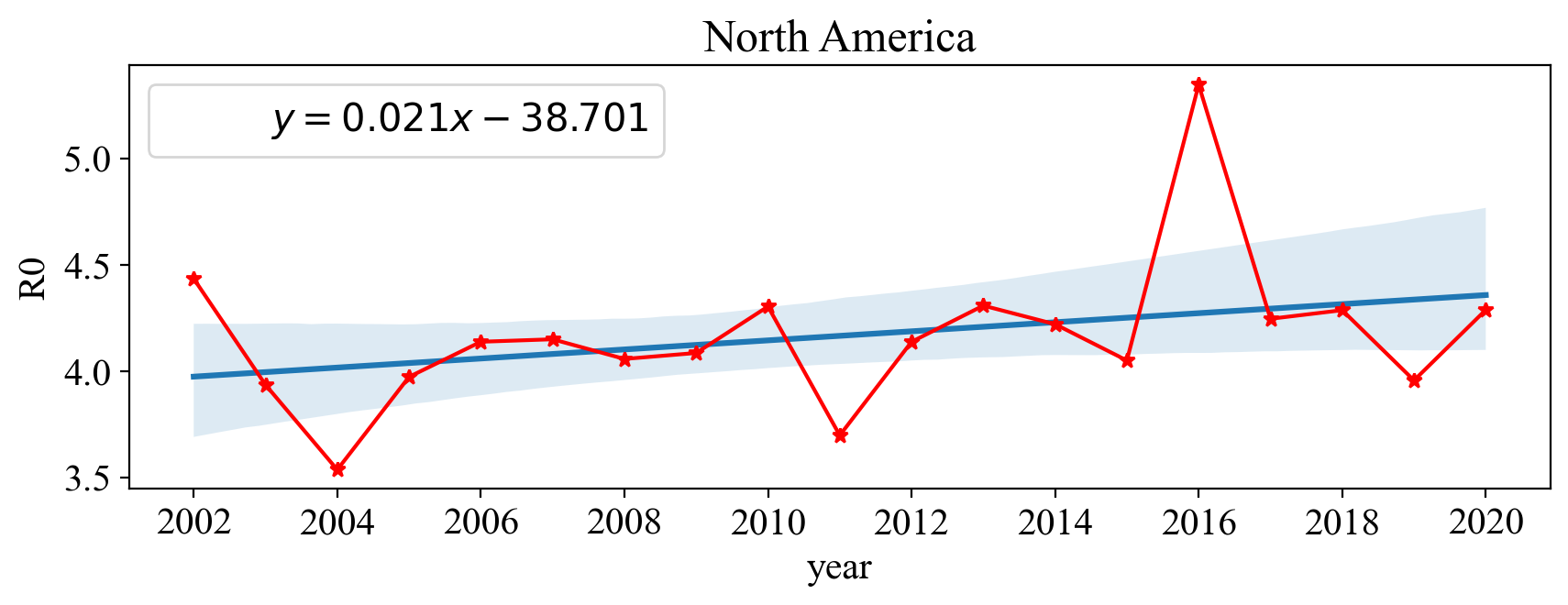}}\\
	\subfloat[Oceania]{\label{} \includegraphics[width=0.45\textwidth]{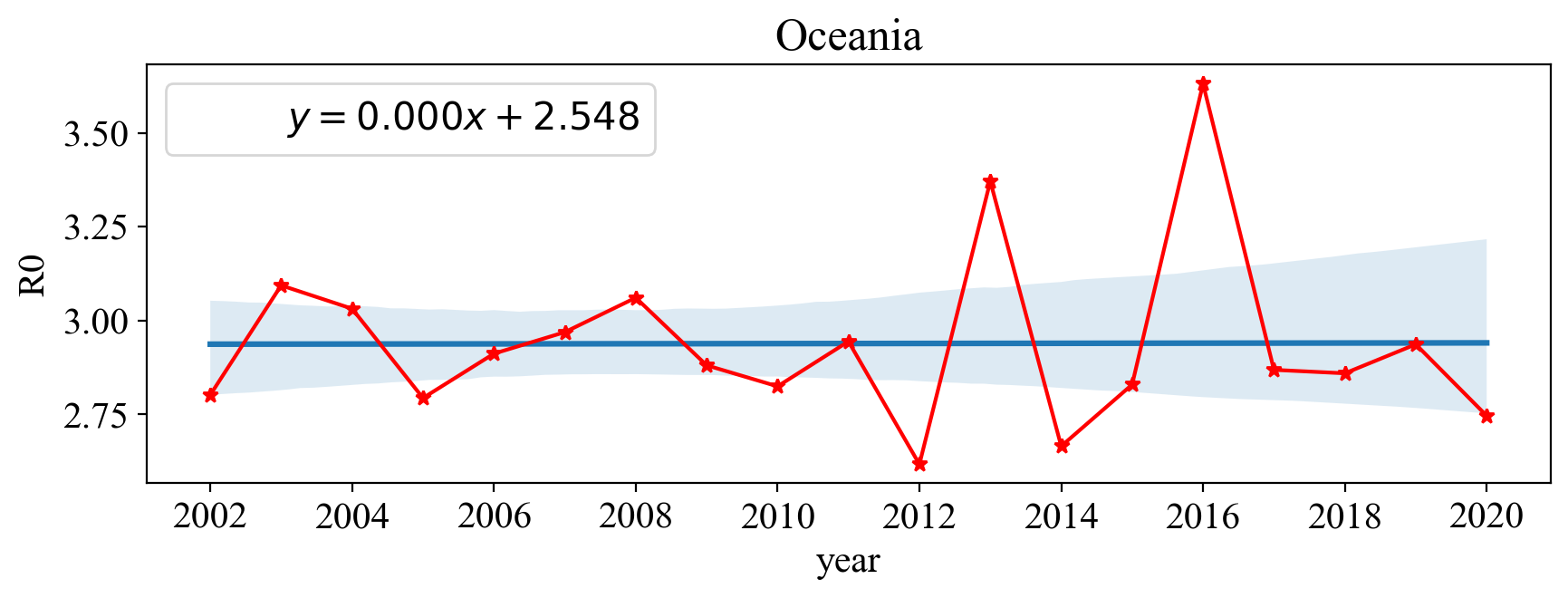}}
	\subfloat[South America]{\label{fig:R0_SA_temporal} \includegraphics[width=0.45\textwidth]{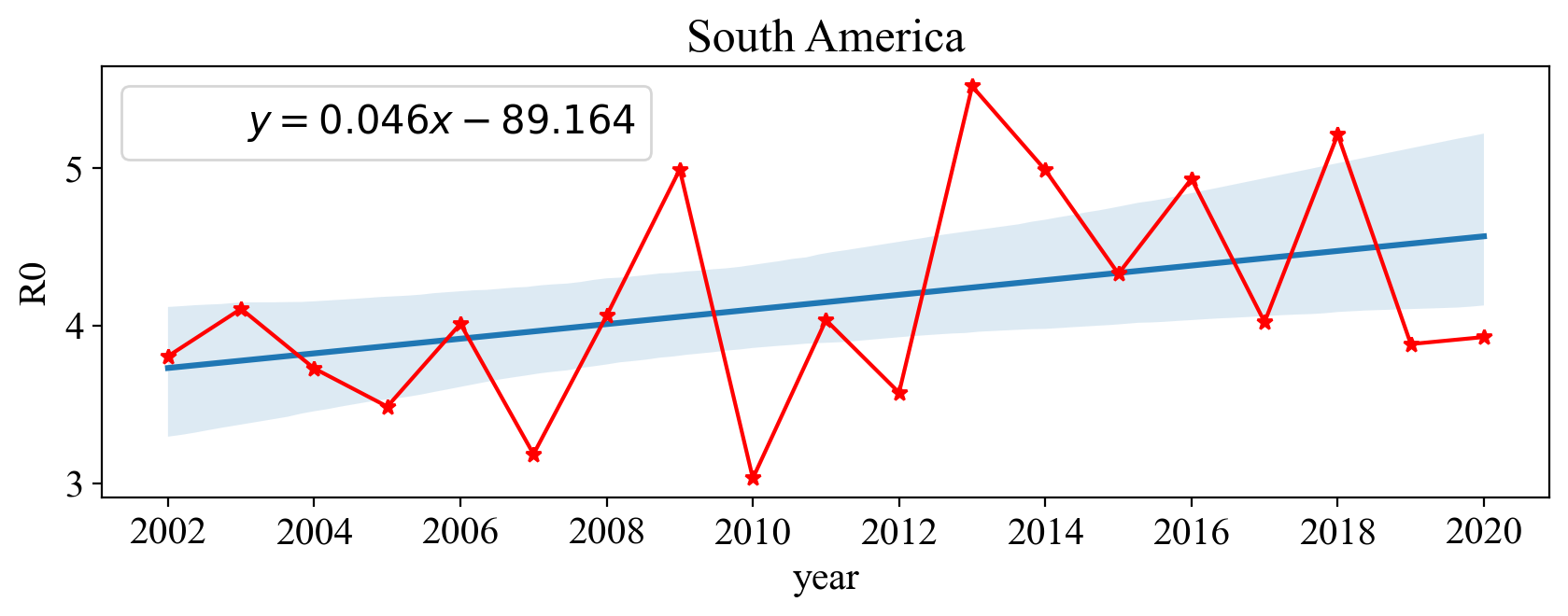}}\\
	\subfloat[Global]{\label{fig:R0_global_temporal} \includegraphics[width=0.45\textwidth]{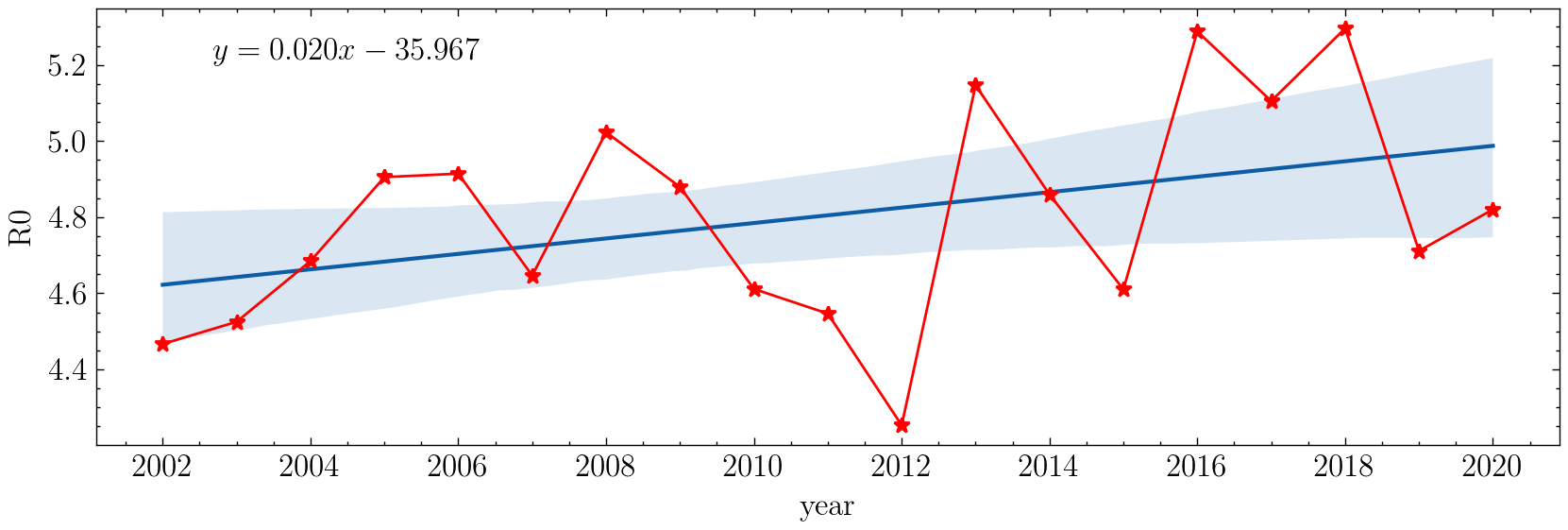}}\\
	\caption{Annual trend of global \rzero for the fire events lasting more than 9 days.}
	\label{fig:R0_temporal}
\end{figure*}

The past twenty years have seen the dramatic evolution of global pattern of $R_0$. Time series of global \rzero in Fig.~\ref{fig:R0_global_temporal} presents an annual trend of global \rzero from 2002 to 2020 for the fire events lasting more than 9 days. Quantitative findings indicate a substantial increase of 8\% in the global \rzero between 2002 (\rzero = 4.47) and 2020 (\rzero = 4.82). The evolution of global \rzero fluctuated substantially but we can still observe a slightly increase by 0.02 annually. With an areal aggregation by continents, from Fig.~\ref{fig:R0_Africa_temporal}-\ref{fig:R0_SA_temporal} and Table~\ref{tab:R0_temporal}, we found a small growth trend of \rzero for most continents over two decades, except for Europe (\rzero decreases by 0.02 per year). To demonstrate a general variation of the fire amount and severity for an area per decade, we also summed up \rzero of all event in each decade and defined a variation ratio, $\rho = (\frac{\sum_{i} R_0^i(2013-2023)}{\sum_{i} R_0^i (2002-2012)} - 1)/100\%$, where $i$ is the index of fire events. Globally, $\rho$ is estimated to be 41\%, means that the sum of \rzero of all events in 2013-2023 has increased by 41\%, compared with that in 2002-2012. Of this total growth, Oceania, North America, and Asia have leading contribution, with the variation ratio per decade are 53\%, 52\%, and 51\%, respectively. 

\begin{figure*}[htbp] 
	\centering
	\subfloat[]{\label{fig:R0_global_1st_deca} \includegraphics[width=0.8\textwidth]{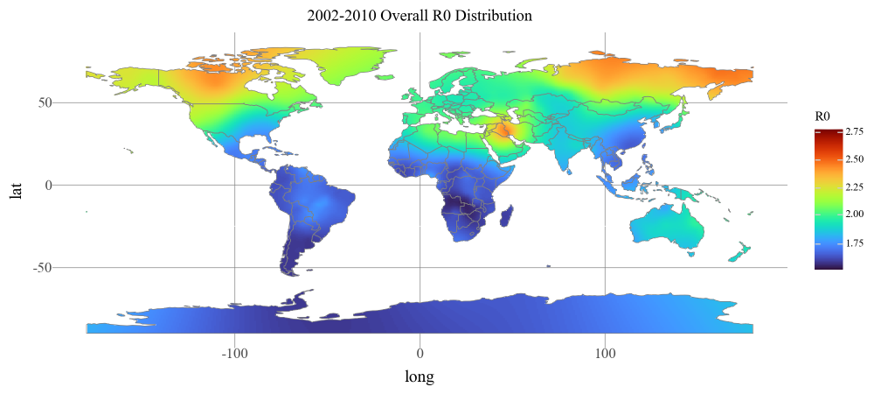}}\\
	\subfloat[]{\label{fig:R0_global_2nd_deca} \includegraphics[width=0.8\textwidth]{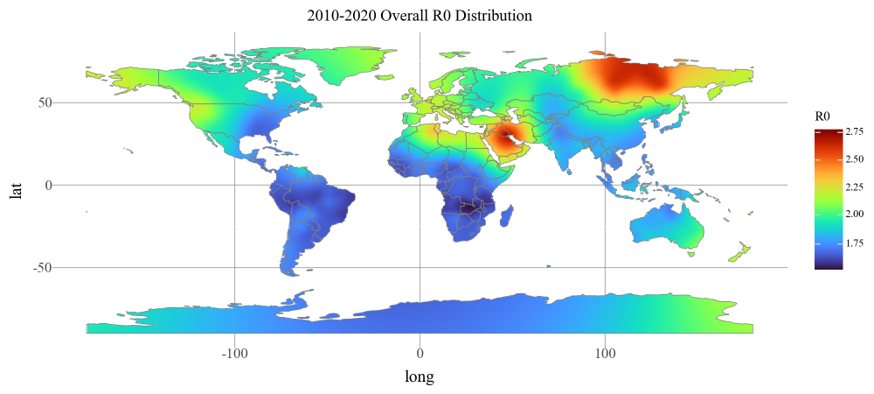}}\\
	\subfloat[]{\label{fig:R0_global_deca_diff} \includegraphics[width=0.8\textwidth]{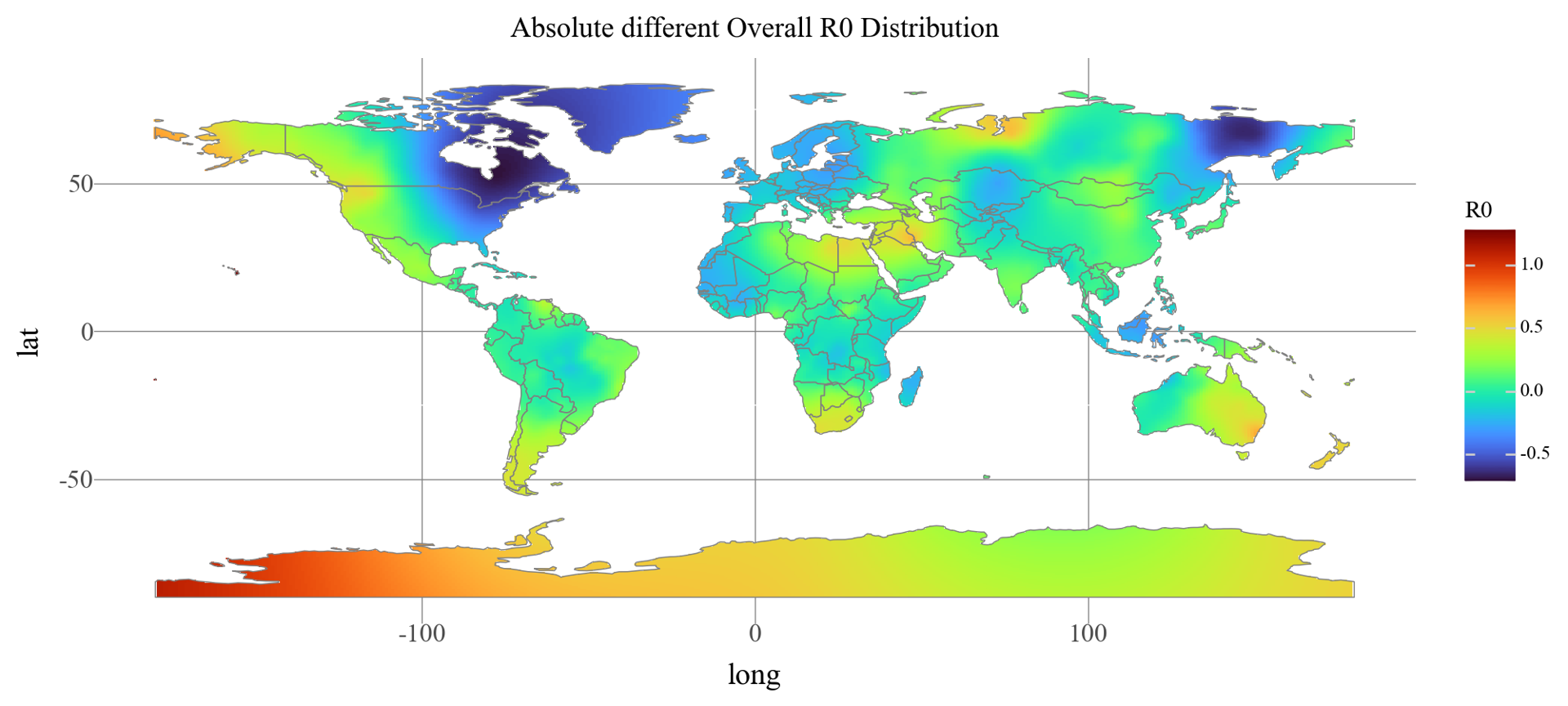}}\\
	\caption{Global distribution of averaged \rzero in the first (a) and second (b) decades of 2002-2020, and the trend during this period (changes per decade). }
	\label{fig:R0_global_two_deca}
\end{figure*}

The global spatial distributions of \rzero in Fig.~\ref{fig:R0_global_1st_deca} and \ref{fig:R0_global_2nd_deca}shows different patterns between the two decades (2002–20012 and 2013–2023),  In Fig.~\ref{fig:R0_global_deca_diff}, from 2002 to 2023, \rzero showed statistically significant increasing trends in Middle East, Western North America, Western Australia, Africa, and South America, whereas notable decreasing trends were found in Eastern North America and Eastern Russia. 

\begin{figure*}[htbp] 
	\centering
	\subfloat[]{\label{} \includegraphics[width=\textwidth]{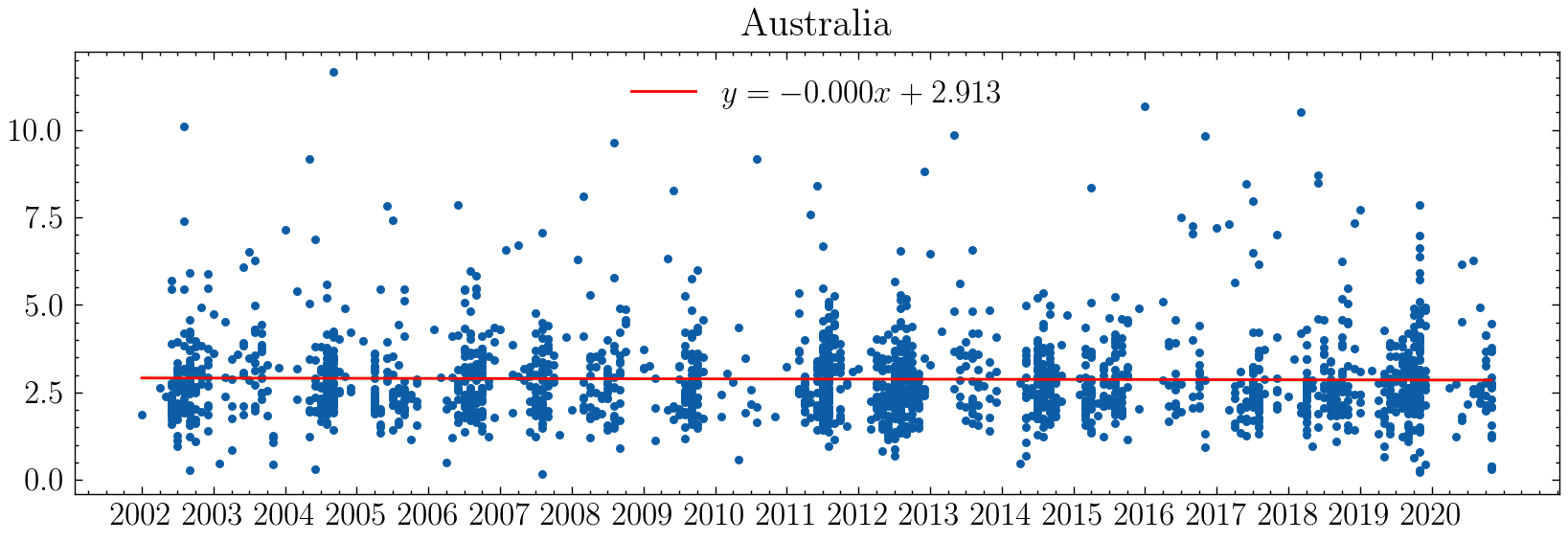}}\\
	\subfloat[]{\label{} \includegraphics[width=\textwidth]{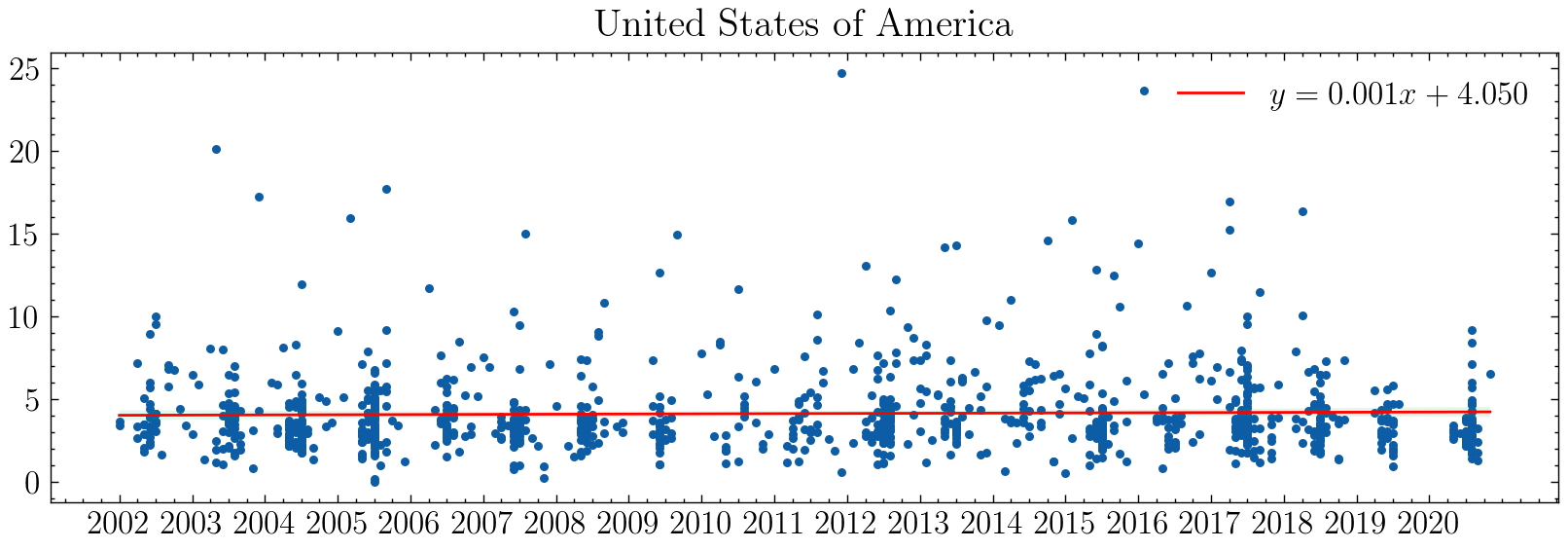}}\\
	\subfloat[]{\label{} \includegraphics[width=\textwidth]{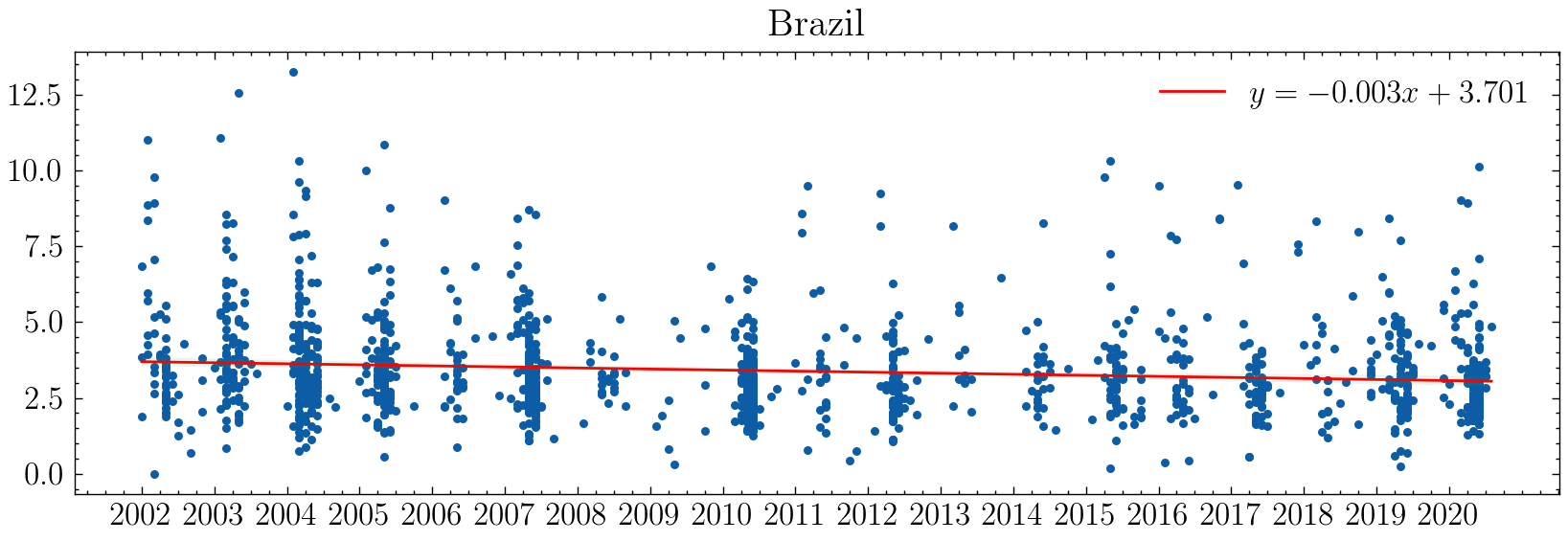}}\\
	\caption{R0 distribution from year 2002 to 2020 in Australia, The United State of America and Brazil, respectively. Fire events are filtered  }
	\label{fig:R0_3countries_temporal}
\end{figure*}

Fig.~\ref{fig:R0_3countries_temporal} shows detailed analyses that are applied to some countries of hotspot, e.g., Australia, The United State of America, and Brazil, where extreme, long-duration wildfires happened.

\begin{table*}[htbp]
	\centering
	\begin{tabular}{ccc}
		\hline
		\hline
		Continent & \rzero Slope & $\rho$ (\%) \\
		\hline
		Africa & 0.04 & 12\\
		Asia & 0.01 & 51\\
		Europe & -0.02 & 29\\
		North America & 0.02 & 52\\
		Oceania & 0.00 & 53\\
		South America & 0.05 & 42\\
		\hline
		Global & 0.02 & 41 \\
		\hline
		\hline
	\end{tabular}
	\caption{$\rho = (\frac{\sum_{i} R_0^i(2013-2023)}{\sum_{i} R_0^i (2002-2012)} - 1)/100\%$, where i is the index of fire events. }
	\label{tab:R0_temporal}
\end{table*}

%

\section*{Discussion}

\section*{Data and methods}\label{sec:data_methods}

\subsection*{Wildfire data}\label{sec:wildfiredata}
We use the novel FireTracks (FT) Scientific Dataset \cite{traxl_2021_4461575} of individual fires. The FT algorithm employs network theory and the individual fires approach to aggregate fire events into spatio-temporal fire clusters that are tracked over space and time. Individual fires are the union of nearest neighbours of active fires in the discrete spacetime grid given by the spatial and temporal resolution of the Moderate Resolution Imaging Spectroradiometer (MODIS) 1-km MOD/MYD14A1 Thermal Anomalies and Fire dataset (Giglio and Justice 2015) that feeds the algorithm. Two fire events are considered neighbours if they are in the same 3-dimensional (latitude, longitude, time) Moore neighbourhood with no spatial or temporal gaps (Fig. 1b). The MOD/MYD14A1 fire product offers an indication of fire activity and has been extensively validated (Morisette et al. 2005; Csiszar et al. 2006; Hawbaker et al. 2008; de Klerk 2008). The collection 6 of the data addresses previous limitations such as frequent false alarms caused by small clearings in the Amazon forests (Friedl et al. 2010), which is particularly helpful for our purpose. The data present low levels of commission errors, but omission errors, which decrease as fire size increases, might occur with fires of short duration, small size or low intensity (Schroeder et al. 2008; Hantson et al. 2013). Also, burnings under dense vegetation cover, heavy smoke or clouds may go undetected (Giglio et al. 2016). 

The FT dataset registers location, time and land cover of individual fires at daily time step, as well as their estimated size, intensity, duration and rate of spread (see Text S1 for the definition of the fire variables). The smallest identifiable fire size and duration is imposed by the spatio-temporal resolution of the MODIS fire data, 0.86 km2 and one day, respectively. Fires with sizes smaller than one fire-data pixel are attributed a size of 0.86 km2 regardless, which may generate some overestimation of burned area. Fires of a single fire-data pixel size (0.86 km2) are not considered for the calculation of rate of spread—the ratio between size and duration. We select those fires within the BLA over the time period from 2002 to 2020 in six land-cover types: croplands, deciduous forests, grasslands, evergreen forests, savannas and woody savannas (see Text S2 and Fig. S2 for the description and spatial distribution of the different land covers, respectively).

The FT algorithm combines the MODIS fire data with land-cover information from the previous year. We use the UMD classification scheme of the 500-m Land Cover Type MCD12Q1 product from MODIS (Sulla Menashe et al. 2019). The collection 6 of the land-cover data includes new gap-filled spectro-temporal features and refinements of the algorithm, which allows a more accurate classification. However, some limitations are known, e.g. grassland areas might be misclassified as savannas, and agriculture can be underrepresented in tropical regions where agricultural fields are small (Friedl et al. 2010). Since the land-cover data has a spatial resolution twice as high as the active fires data (0.21 vs. 0.86 km2), the FT algorithm associates four values of land cover with every MODIS active fire within a particular individual fire. Fires are assigned a dominant land cover when at least 80\% of all the land-cover values within them belong to the same land-cover type. Fires that do not fulfil this criterion are discarded from the analysis. In this way, we ensure that the FT’s fire characteristics estimated for each land-cover type are not a combination of values from different land covers.

We employ the GFA dataset (Andela et al. 2019b), the most extensive study on individual fires covering the BLA so far, to perform a comparison of our estimated fire characteristics. The GFA is derived from the MODIS collection 6 500-m Burned Area MCD64A1 product (Giglio et al. 2018) and spans from 2003 to 2016. The quality of the algorithm, as for the FT’s, highly depends on the inherent limitations of the data that serve as input. Fires of 0.21 km2—the smallest identifiable fire size—are not taken into account when calculating rate of spread. The GFA algorithm tracks the daily progression of individual fires to produce a set of metrics on fire behaviour such as fire size, duration, daily expansion, fire line length, speed and direction of spread. We select from the FT dataset the fires identified in the BLA over the period 2003–2016—the same 14-year time window when data from the GFA is available—and compare fire size, duration and rate of spread—the variables present in both datasets—occurring in croplands, forests, grasslands and savannas (Text S2).


\subsection*{Methods}\label{sec:methods}
\subsubsection*{Wildfire dynamics}
To well model the spread and describe the dynamics of each fire event, we use a well-known compartment model called susceptible-infected-recovered (SIR) with susceptible ($S$), Infectious ($I$), and Removed ($R$) compartments. Within the perimeter of an individual fire, there are N spatio-temporal fire grids in total. Susceptible spatio-temporal grids are flammable but have never been burned yet, while they can get ignited through contact with a primary fire event. Burning grids are categorized into Infectious state. Then, fires will get extinguished or go out after a period of time, meaning Infectious s will transition into the Removed state. The removed state represents fires can no longer resurge once gone out. 

Thus, analogous to the modeling of infectious disease spread in a population, the dynamics of an individual fire $i$ can be described by SIR model: 
\begin{align*}
	\frac{dS_i(t)}{dt} =& -\beta S_i(t) I_i(t)\\
	\frac{dI_i(t)}{dt} =& \beta S_i(t) I_i(t) - \mu I_i(t)\\
	\frac{dR_i(t)}{dt} =& \mu I_i(t)
\end{align*}
where $\beta$ represents the transmission rate of an individual fire event, and $\mu$ represents the dissipation rate of the fire. $\beta \in [0,1]$ and $\mu \in [\frac{1}{d},1]$ where $d$ is the duration of a fire event. We introduce a basic reproduction number, $R_0 = \frac{\beta S(0)}{\mu}$, which represents the average number of secondary infections produced by a single infected individual in a completely susceptible population (i.e. $S(t) = S(0)$). Fitting the empirical data with the SIR model, one can obtain an evaluable criteria, $R_0$, according to which the government or managers can appraise the fire risk of jurisdictions.

\subsubsection*{Learning the relationship between \rzero and land-cover types}
Conventionally, fire regimes (or behaviors) is characterized by a few parameters \cite{fire1010009}: fire size, duration, fireline intensity (FLI), Rate of Spread (ROS), spotting, flame length (FL), etc. For example, Ana Cano‑Crespo et al. \cite{cano2023characterization} analyzed fire regimes in six different land-cover classes. Fires in savannas and evergreen forests were found to burn the largest areas and are the most long lasting. Inspired by their analysis, in this paper, the correlation between $R_0$ and different land-cover types is studied. Rather than assuming a specific statistical or functional relationship between $R_0$ and land-cover types, we learn it directly from the data. Specifically, we use neural networks and random forests to learn the potentially nonlinear relationship in ways that are highly conditional on the state of other environmental variables.


\nocite{o2021age}
\nocite{chang2021mobility}
\nocite{goldstein2020demographic}
\nocite{nytdata}
\nocite{ew-unitedway}
\nocite{worldbanklabor}
\nocite{world2020sage}
\nocite{khubchandani2021covid}
\nocite{vac_demo}
\nocite{illsley1987measurement}
\nocite{leclerc1990differential}
\nocite{berndt2003measuring}
\nocite{dixon1987bootstrapping}
\nocite{delaware-list}
\nocite{glaeser2020jue}
\nocite{sanchez2020jobs}
\nocite{dudley2020disparities}
\nocite{household_size}
\nocite{svi2018}
\nocite{tzeng2011multiple}


\printbibliography[title={References and Notes}]

@dataset{traxl_2021_4461575,
	author       = {Traxl, Dominik},
	title        = {The FireTracks Scientific Dataset},
	month        = apr,
	year         = 2021,
	publisher    = {Zenodo},
	version      = {1.0.0},
	doi          = {10.5281/zenodo.4461575},
	url          = {https://doi.org/10.5281/zenodo.4461575}
}

@article{cano2023characterization,
  title={Characterization of land cover-specific fire regimes in the Brazilian Amazon},
  author={Cano-Crespo, Ana and Traxl, Dominik and Prat-Ortega, Gen{\'\i}s and Rolinski, Susanne and Thonicke, Kirsten},
  journal={Regional Environmental Change},
  volume={23},
  number={1},
  pages={19},
  year={2023},
  publisher={Springer}
}

@Article{fire1010009,
AUTHOR = {Tedim, Fantina and Leone, Vittorio and Amraoui, Malik and Bouillon, Christophe and Coughlan, Michael R. and Delogu, Giuseppe M. and Fernandes, Paulo M. and Ferreira, Carmen and McCaffrey, Sarah and McGee, Tara K. and Parente, Joana and Paton, Douglas and Pereira, Mário G. and Ribeiro, Luís M. and Viegas, Domingos X. and Xanthopoulos, Gavriil},
TITLE = {Defining Extreme Wildfire Events: Difficulties, Challenges, and Impacts},
JOURNAL = {Fire},
VOLUME = {1},
YEAR = {2018},
NUMBER = {1},
ARTICLE-NUMBER = {9},
URL = {https://www.mdpi.com/2571-6255/1/1/9},
ISSN = {2571-6255},
DOI = {10.3390/fire1010009}
}

@Article{fire1010018,
AUTHOR = {Nauslar, Nicholas J. and Abatzoglou, John T. and Marsh, Patrick T.},
TITLE = {The 2017 North Bay and Southern California Fires: A Case Study},
JOURNAL = {Fire},
VOLUME = {1},
YEAR = {2018},
NUMBER = {1},
ARTICLE-NUMBER = {18},
URL = {https://www.mdpi.com/2571-6255/1/1/18},
ISSN = {2571-6255},
DOI = {10.3390/fire1010018}
}

@article{boer2020unprecedented,
  title={Unprecedented burn area of Australian mega forest fires},
  author={Boer, Matthias M and Resco de Dios, V{\'\i}ctor and Bradstock, Ross A},
  journal={Nature Climate Change},
  volume={10},
  number={3},
  pages={171--172},
  year={2020},
  publisher={Nature Publishing Group UK London}
}

@article{o2021age,
  title={Age-specific mortality and immunity patterns of SARS-CoV-2},
  author={O’Driscoll, Megan and Dos Santos, Gabriel Ribeiro and Wang, Lin and Cummings, Derek AT and Azman, Andrew S and Paireau, Juliette and Fontanet, Arnaud and Cauchemez, Simon and Salje, Henrik},
  journal={Nature},
  volume={590},
  number={7844},
  pages={140--145},
  year={2021},
  publisher={Nature Publishing Group}
}

@article{chang2021mobility,
  title={Mobility network models of COVID-19 explain inequities and inform reopening},
  author={Chang, Serina and Pierson, Emma and Koh, Pang Wei and Gerardin, Jaline and Redbird, Beth and Grusky, David and Leskovec, Jure},
  journal={Nature},
  volume={589},
  number={7840},
  pages={82--87},
  year={2021},
  publisher={Nature Publishing Group}
}

@techreport{world2020sage,
  title={WHO SAGE roadmap for prioritizing the use of COVID-19 vaccines in the context of limited supply: an approach to inform planning and subsequent recommendations based upon epidemiologic setting and vaccine supply scenarios, 13 November 2020},
  %author={World Health Organization},
  year={2020},
  institution={World Health Organization}
}

@Misc{delaware-list,
howpublished = {\url{https://www.coastalpoint.com/news/coronavirus/delaware-list-of-essential-and-non-essential-businesses/pdf_b43d669c-6c8f-11ea-bde7-971f96836402.html}},
%note = {Accessed~on~July~18th,~2021},
title = {Delaware List of Essential and Non-Essential Businesses.},
}

@article{glaeser2020jue,
  title={JUE insight: How much does COVID-19 increase with mobility? Evidence from New York and four other US cities},
  author={Glaeser, Edward L and Gorback, Caitlin and Redding, Stephen J},
  journal={Journal of urban economics},
  pages={103292},
  year={2020},
  publisher={Elsevier}
}

@article{dudley2020disparities,
  title={Disparities in age-specific morbidity and mortality from SARS-CoV-2 in China and the Republic of Korea},
  author={Dudley, Joseph P and Lee, Nam Taek},
  journal={Clinical Infectious Diseases},
  volume={71},
  number={15},
  pages={863--865},
  year={2020},
  publisher={Oxford University Press US}
}

@Misc{household_size,
howpublished = {\url{https://www.statista.com/statistics/242189/disitribution-of-households-in-the-us-by-household-size/}},
%note = {Accessed~on~July~30th,~2021},
title = {US Census Bureau. (December 1, 2020). Distribution of households in the United States from 1970 to 2020, by household size.},
}

@article{goldstein2020demographic,
  title={Demographic perspectives on the mortality of COVID-19 and other epidemics},
  author={Goldstein, Joshua R and Lee, Ronald D},
  journal={Proceedings of the National Academy of Sciences},
  volume={117},
  number={36},
  pages={22035--22041},
  year={2020},
  publisher={National Acad Sciences}
}

@book{tzeng2011multiple,
  title={Multiple attribute decision making: methods and applications},
  author={Tzeng, Gwo-Hshiung and Huang, Jih-Jeng},
  year={2011},
  publisher={CRC press}
}

@article{leclerc1990differential,
  title={Differential mortality: some comparisons between England and Wales, Finland and France, based on inequality measures},
  author={Leclerc, Annette and Lert, France and FABIEN, C{\'E}CILE},
  journal={International journal of epidemiology},
  volume={19},
  number={4},
  pages={1001--1010},
  year={1990},
  publisher={Oxford University Press}
}

@incollection{illsley1987measurement,
%  title={The measurement of inequality in health},
%  author={Illsley, Raymond and Le Grand, Julian},
%  booktitle={Health and economics},
%  pages={12--36},
%  year={1987},
%  publisher={Springer}
%}

@inproceedings{berndt2003measuring,
  title={Measuring healthcare inequities using the Gini index},
  author={Berndt, Donald J and Fisher, John W and Rajendrababu, Rama V and Studnicki, James},
  booktitle={Proceedings of the 36th Annual Hawaii International Conference on System Sciences, 2003.},
  pages={10--pp},
  year={2003},
  organization={IEEE}
}

@Misc{nytdata,
howpublished = {\url{https://github.com/nytimes/covid-19-data}},
%note = {Accessed~on~Feb~6th,~2021},
title = {Coronavirus (Covid-19) Data in the United States.},
author = {The~New~York~Times}
}

@Misc{ew-unitedway,
howpublished = {\url{https://unitedwaynca.org/stories/us-states-essential-workers/}},
%note = {Accessed~on~Aug~23th,~2021},
title = {US States with the most essential workers.},
}

@Misc{worldbanklabor,
howpublished = {\url{https://data.worldbank.org/indicator/SL.TLF.TOTL.IN?locations=US}},
%note = {Accessed~on~Aug~23th,~2021},
title = {Labor force in the United States.},
}

@article{dixon1987bootstrapping,
  title={Bootstrapping the Gini coefficient of inequality},
  author={Dixon, Philip M and Weiner, Jacob and Mitchell-Olds, Thomas and Woodley, Robert},
  journal={Ecology},
  volume={68},
  number={5},
  pages={1548--1551},
  year={1987},
  publisher={JSTOR}
}

@Misc{svi2018,
howpublished = {\url{https://www.atsdr.cdc.gov/placeandhealth/svi/data_documentation_download.html}},
title = {CDC/ATSDR Social Vulnerability Index 2018.},
organization={Centers for Disease Control and Prevention and Agency for Toxic Substances and Disease Registry/Geospatial Research, Analysis, and Services Program},
%note = {Accessed~on~Sept~15th,~2021},
}

@article{sanchez2020jobs,
  title={Which jobs are most vulnerable to COVID-19? What an analysis of the European Union reveals},
  author={Sanchez, Daniel Garrote and Parra, Nicolas Gomez and Ozden, Caglar and Rijkers, Bob},
  journal={World Bank Research and Policy Briefs},
  number={148384},
  year={2020}
}

@article{khubchandani2021covid,
  title={COVID-19 vaccination hesitancy in the United States: a rapid national assessment},
  author={Khubchandani, Jagdish and Sharma, Sushil and Price, James H and Wiblishauser, Michael J and Sharma, Manoj and Webb, Fern J},
  journal={Journal of Community Health},
  volume={46},
  number={2},
  pages={270--277},
  year={2021},
  publisher={Springer}
}

@Misc{vac_demo,
%howpublished = {\url{https://covid.cdc.gov/covid-data-tracker/\#\#vaccination-demographic}},
%title = {COVID Data Tracker, Vaccination Demographics.},
%organization={Centers for Disease Control and Prevention, U.S.},
%%note = {Accessed~on~Sept~15th,~2021},
%}

\section*{Acknowledgments}
\textbf{Funding:}
\textbf{Author Contributions:}
\textbf{Competing interests:}
\textbf{Data and materials availability:}

\section*{Supplementary materials}
Materials and Methods\\
Figs S1 to S17\\
Tables S1 to S14\\
References (58-69)


\clearpage


\newpage

\clearpage
\begin{spacing}{1.0}
\noindent\textbf{Figure 2: Social utility and equity of vaccine distribution strategies that prioritize disadvantaged communities under a single demographic dimension.}
\textbf{(a)} Changes in social utility and equity, compared to the \textit{Homogeneous} baseline.
The red/blue points respectively represent the strategies prioritizing the most/least disadvantaged communities.
The $1^{st}$ to $4^{th}$ quadrant respectively represents: (i) simultaneously improving utility and equity, (ii) improving equity but damaging utility, (iii) improving utility but damaging equity, and (iv) simultaneously damaging utility and equity.
\textbf{(b)} Changes in three dimensions of equity, compared to the \textit{Homogeneous} baseline.
Highlighted grids indicate degradation in the corresponding dimension of equity.
\textbf{(c)} 
Social utility under different scenarios of vaccine hesitancy.
When vaccine hesitancy in low-income communities is stronger, the benefit on social utility brought by prioritizing disadvantaged communities diminishes.
In extreme scenarios, the benefit of prioritizing disadvantaged communities characterized by income is completely erased, making it inferior to the baseline. 
\textbf{(d)} 
Joint probability distribution of demographic features, where brighter colors indicate larger probability density.
The correlations between (i) the percentage of older adults and the average household income, (ii) the percentage of older adults and the percentage of essential workers, (iii) the average household income and the percentage of essential workers are (i) $+0.14$, (ii) $-0.2$, and (iii) $+0.28$, respectively, displaying the mismatch among disadvantaged communities in different demographic dimensions.
\end{spacing}

\end{document}